\documentstyle[twocolumn,aps]{revtex}

\begin{document}
\input{psbox.tex}
\draft

\title{
Energy gap in superconducting fullerides: optical and tunneling studies.  
}

\author{Daniel Koller, Michael C. Martin and L\'aszl\'o Mih\'aly}
\address{Department of Physics, SUNY at Stony Brook, Stony Brook, NY 
11794-3800}
\author{Gy\"orgy Mih\'aly\cite{str}, G\'abor Oszl\'anyi\cite{str2}, 
Gabriel Baumgartner and L\'aszl\'o Forr\'o}
\address{Department of Physics, Ecole Polytechnique Federale de Lausanne 
CH-1015 Lausanne, Switzerland}

\date{\today}
\maketitle

\begin{abstract}
Tunneling and optical transmission studies have been performed on 
superconducting samples of Rb$_{3}$C$_{60}$.  At temperatures much below the 
superconducting transition temperature $T_{\rm c}$ the energy gap is 
$2\Delta=5.2\pm 0.2$meV, corresponding to $2\Delta/k_B T_{\rm c} = 4.2$.  The 
low temperature density of states, and the temperature dependence of the 
optical conductivity resembles the BCS behavior, although there is 
an enhanced ``normal state" contribution.  The results indicate that this
fulleride material is an s-wave superconductor, but the superconductivity 
cannot be described in the weak coupling limit.  
\end{abstract}
\pacs{PACS:  74.70.Wz 74.25.Gz 74.50.+r}
\narrowtext

The symmetry of the order parameter and the magnitude of the energy gap are 
two of the most fundamental properties of superconductors.  While there is 
a general agreement about the {\it s}-wave nature of superconductivity in the 
alkali metal fulleride compounds, the magnitude of the energy gap is not known
with sufficient accuracy.

Why is the determination of the energy gap so important? Let us consider, 
for example, the Eliashberg theory of phonon-mediated superconductivity, and 
represent the phonons by a single Einstein mode.\cite{allen,carbotte}  
There are two parameters in this theory:  The  phonon
frequency $\omega_0$ and the coupling between electrons and phonons 
$\lambda$.
(The latter one may be viewed as an ``effective" electron -- phonon coupling 
constant, incorporating the ``true" $\lambda$ and the electron -- electron 
repulsion parameter $\mu$.\cite{allen,mcmillan})  In the weak coupling (BCS) 
limit the  quantity $\omega_0 \exp (-1/\lambda)$ is the only relevant 
combination of parameters.  In that limit the
ratio of the low temperature value of the superconducting energy 
gap to the critical temperature is always $2\Delta/k_B T_{\rm c} = 
3.53$.  If the coupling is strong, however, knowledge of the 
critical temperature and the  low temperature gap provides us with $\lambda$ 
and $\omega_0$. Once the magnitudes of these parameters are known, 
theoretical models for the microscopic mechanism of 
superconductivity \cite{gunnarson} can be placed in the proper context.  

A review of the literature reveals a great disparity 
between the various measurements of the superconducting energy gap
in Rb$_3$C$_{60}$.  Direct spectroscopic 
methods to determine the gap include optical spectroscopy, tunneling and 
photoemission. The first optical study by Rotter {\it et al.} \cite{rotter} 
resulted in values of $\eta \equiv 2\Delta/k_B T_{\rm c} \approx 3-4$.  
Fitzgerald {\it et al.} \cite{fitzgerald} obtained  
$\eta \approx 3-5$ from transmission measurements on thin film  samples. In 
more recent reflectance measurements on polycrystalline\cite{degiorgipoly} 
and single crystal\cite{degiorgisingle} samples, Degiorgi {\it et al.} obtained 
$\eta \approx 2.98$ and $\eta \approx 3.45$, respectively.   Zhang {\it et al.} \cite{zhang} 
reported tunneling measurements yielding a value greater than 5, whereas 
Jess and co-workers measured values between 2 and 4 in an STM study 
\cite{jess}. Photoemission experiments by the Argonne-University of Illinois 
collaboration \cite{gu} led to $\eta \approx 4.1 \pm 0.4$.  Nuclear magnetic 
resonance \cite{tycko,stenger}  and muon spin relaxation \cite{kiefl} yielded 
values compatible with the BCS weak coupling limit. (All numerical values 
summarized here apply to Rb$_{3}$C$_{60}$.)

An accurate determination of $\eta$, and the detailed fit of the 
theoretical curves to the experimental results answers a fundamental 
question: Is the coupling between the electrons weak enough for 
the BCS theory to describe superconductivity?
In this Letter we report optical transmission and tunneling measurements
unambiguously demonstrating that the ratio $2 \Delta /k_B T_{\rm c}$ is $4.2\pm 0.2$, 
significantly larger than the BCS value. Since the weak coupling conditions
are not satisfied, we are able to deduce the 
phonon coupling constant and the characteristic phonon frequency, obtaining 
$\lambda=0.9$ and $\omega_0=210$cm$^{-1}$ ($\hbar \omega_0=26$meV). 
The electron -- phonon coupling is consistent with the value determined from DC 
resistivity measurements \cite{vareka}; the phonon frequency is close to the
lowest lying Raman frequency of the C$_{60}$ molecule \cite{martin}.

\begin{figure}
%Shift caption closer to real bottom of figure. 
%This is done by chopping off x cm from the bottom.
\psyoffset=-4.7cm
\pshtincr=-4.7cm
%Add right margin
\pswdincr=1cm
%Scale figure to the right horizontal size
\psboxto(3.4in;0in){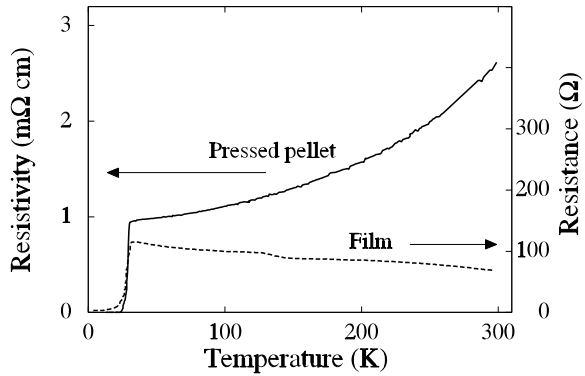}
\caption{Resistivity and resistance of the thin film and pressed pellet
samples used in the optical and tunneling measurements, respectively. 
%The 
%resistivity of the thin film sample at room temperature is about 
%10m$\Omega$cm
}  
%\caption{Resistivity vs. temperature for the pressed pellet sample of 
%Rb$_{3}$C$_{60}$ used in the tunneling measurements.}  
\label{resist}
\end{figure}

The pressed pellet 
samples for the tunneling measurements were made by heating and mixing
stoichiometric amounts of Rb and C$_{60}$ in a quartz tube for several weeks. 
%\cite{forrosample}
According to powder X-ray diffraction, the sample contained
no secondary phases.   
The material was pressed into a pellet, with four gold wires embedded for
electrical contacts.  The resistivity as a function 
of the temperature is shown in Figure \ref{resist}.  The 
resistivity of the pressed pellet sample was as low as the resistivity of 
single crystals\cite{vareka}.  The superconducting transition temperature was 
30K.

The optical measurements were performed on samples prepared by deposition of 
a thin C$_{60}$ film on  a Si substrate and subsequent exposure to Rb vapor in a 
sample cell mounted in a Bomem MB-155 spectrometer.\cite{koller}  The composition of the 
sample was monitored by observing the $F_{1u}(4)$ vibrational mode of the C$_{60}$ 
molecule \cite{mike}. The thickness of the film was $1.0\pm0.1\mu$m. 
Scanning force microscope images
of the sample surface showed an irregular surface with poorly matching crystal 
faces.  
From room temperature to 30K there was a factor of two increase 
of resistance, typical of thin film samples \cite{bell} and indicative of 
granular morphology (Figure \ref{resist}).  
The superconducting transition occurred at 30K; a 
smearing of the transition suggested a distribution of critical 
temperatures, also consistent with granular morphology. 

\begin{figure}
%Shift caption lower from bottom of figure. 
%This is done by adding x cm to the bottom.
\psyoffset=1cm
\pshtincr=1cm
%Chop off right side
\pswdincr=-2cm
%Scale figure to the right horizontal size
\psboxto(3.4in;0in){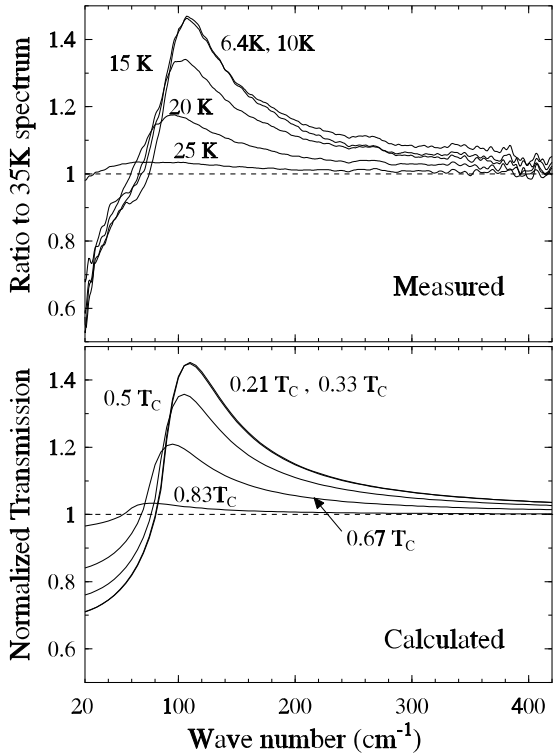}
\caption{
(top) Far infrared transmission of the sample at 
temperatures below the superconducting transition.  The curves are  
normalized to the transmission at $T = 35K$ in order to eliminate the  
features due to the silicon substrate. (bottom) Calculated transmission curves, 
based on dirty limit BCS results. 
}
\label{transmiss}
\end{figure}

The far--IR transmission spectra were measured as a function of 
temperature at the U4IR beamline of the National Synchrotron Light Source, 
at Brookhaven National Laboratory. The sample cell was mounted on a 
Helitran He flow refrigerator and the spectra were taken in a Nicolet 
20F Rapid Scan FTIR spectrometer.  Due to the capacitive inter-grain coupling,  
AC measurements (including the optical transmission) are
expected to be less sensitive to the grain boundaries then the DC transport.  
In accordance with this expectation, the optical transmission of the film  
decreased upon cooling, corresponding to the increasing intrinsic 
conductivity of the crystallites.  

The upper panel in Figure \ref{transmiss} shows the low temperature transmission
results, normalized to the spectrum taken in the normal state at 35K.  The 
pronounced peak in the optical transmission is a direct 
evidence for the sharp energy gap in the real part of the optical 
conductivity at $\hbar \omega = 2 \Delta$.  This feature was first 
observed for superconductors, and discussed in 
detail, by Glover and Tinkham \cite{tinkham}.  

\begin{figure}
\psyoffset=-4cm
\pshtincr=-4cm
\psboxto(3.5in;0in){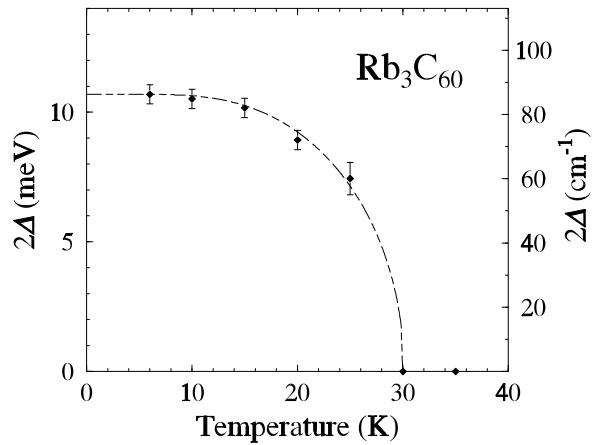}
\caption{Temperature dependence of the energy gap evaluated from the 
optical transmission data. The dashed line represents the BCS temperature 
dependence, scaled to satisfy $2\Delta=4.1k_B T_{\rm c}$.  
}
\label{gaptemp}
\end{figure}

Calculated transmission curves are shown in the lower panel of Figure 
\ref{transmiss}.  In the absence of readily available Eliashberg 
results for the density of states and optical conductivity, we 
modified the BCS result\cite{mattis} to include the effects of
inelastic pair breaking\cite{rainer} in a phenomenological way.  
The total optical transmission $t$ was represented in a two component model, 
in the form of $t=w t_{sc}+ (1-w) t_{n}$, 
where $t_{n}$ is the transmissions of a Drude metal, and $t_{sc}$, was 
calculated from the Mattis-Bardeen\cite{mattis} conductivity
of a superconductor.  This approach also accounts for the inevitable   
imperfections in the sample. In the best fits the 
weight factor was about $w=30$\% at the lowest 
temperature, and it decreased as the temperature approached $T_{\rm c}$.  
The 
energy gap as a function of the temperature, evaluated from these fits, is 
shown in Figure \ref{gaptemp}.  Notice that the energy gap at zero 
temperature corresponds to $\eta \approx 4.1 \pm 0.2$.  
For comparison, the temperature dependence of the weak 
coupling (BCS) gap, scaled to $2\Delta=4.1k_B T_{\rm c}$, 
is also shown in the Figure.

\begin{figure}
\psyoffset=-2cm
\pshtincr=-2cm
\psboxto(3.4in;0in){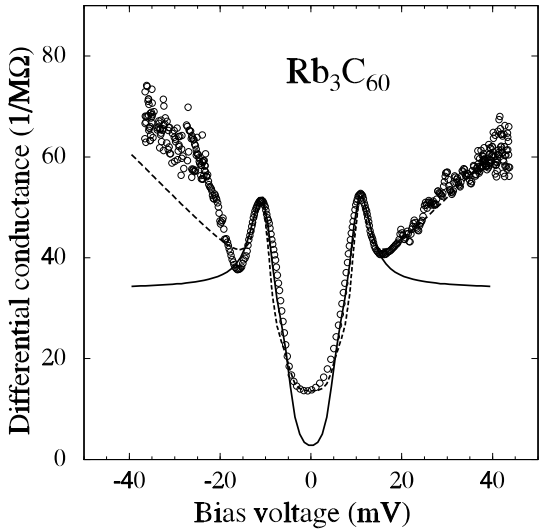}
\caption{Differential conductance of the break junction made of a 
polycrystalline RbC$_{60}$ pellet (open circles).  The continuous line is a three 
parameter fit using a modified BCS formula for the density of states (see 
text).  The dashed line was obtained by including an additional leakage 
conductance at zero bias,  and the voltage dependence of the transmission 
function of the junction.  
}
\label{tunnel}
\end{figure}

For the tunneling measurements
the pressed pellets were first cooled to low temperature.  A break junction was 
made and controlled by mechanical means.  The device was similar 
to the one used in the tunneling studies of high $T_{\rm c}$ superconductors 
\cite{break}.  The sample was in He atmosphere to 
prevent the degradation of the material due to the chemical reaction with 
oxygen.  The differential conductance of a typical junction at 
$T = 4.2K$ is shown in Figure \ref{tunnel}.  The poor thermal stability 
of the junction prevented us from measuring the temperature dependence of 
the tunneling.

Ideal break junctions are superconductor -- insulator -- superconductor (SIS) 
junctions. The tunneling current is described by
$I \propto \int gfg'f' {\rm d} E \;$
where $g=g(E)$ and $g'=g(E-eV)$ are the densities of states and 
$f=f(E)$ and $f'=f(E-eV)$ are Fermi functions 
\cite{tunneltheory}. The simple BCS density of 
states is often modified to 
\begin{equation}
g(E) \approx  \left\vert  { {\rm Re} { E+ {\rm i} \delta \over \sqrt{(E+{\rm i} \delta)^2 + \Delta^2}} } \right\vert \;,
\label{denstate}
\end{equation}
where the phenomenological parameter $\delta$ accounts for the pair breaking
effects\cite{rainer,dynes}.  The continuous line in 
Figure \ref{tunnel} was obtained by adjusting $\Delta$, $\delta$ and the 
vertical scale factor of the differential conductance.  Apart from the 
elevated zero bias conductance and the upturn of conductivity at larger 
bias voltages, the agreement is satisfactory.  (The parameters are 
$2\Delta \approx 5.2$meV  and $\delta \approx 1.5$meV).  A more complete calculation, 
including an additional leakage conductance at zero bias, and a voltage 
dependent tunneling rate \cite{wolf}, results in an even better fit 
(dashed line).  
Most importantly, the Figure demonstrates that the value of the energy gap  is 
not sensitive to the processing of the data:  the peak position of the raw 
data, and the two fitting procedures yield $2\Delta \approx 5.2$meV, corresponding 
to $\eta \approx 4.3 \pm 0.2$

The accuracy of the results and the agreement between the two principal 
spectroscopic methods (optics and tunneling) leave little doubt about the 
magnitude of the low temperate energy gap in Rb$_{3}$C$_{60}$.  The ratio  
$\eta \equiv 2\Delta / k_B T_{\rm c} \approx 4.2 \pm 0.2$ is obtained, which is clearly 
beyond the weak coupling value. 
%If the weak coupling conditions are not satisfied, then some of the 
%assumptions we made during the evaluation of the experimental data become 
%questionable.  For example, 
%the density of states is not described by equation \ref{denstate}, the 
%Mattis-Bardeen result for the optical conductivity is replaced by a more 
%complicated function, the ``generic" $\Delta(T)$ function used in Figure 
%\ref{gaptemp} is not applicable, etc.  Nevertheless,  the most important 
%features  in these functions still closely resemble to the BCS result (the 
%density of states is still zero over an energy range of $\pm \Delta$, the 
%optical conductivity has a sharp onset at $\pm 2\Delta$,  the gap is a 
%smoothly decreasing function of the temperature, etc.).   For $\eta \approx 4.2$ 
%we can still use the BCS functions in our fits, except for the BCS value 
%for the gap.  
This result is compatible with, although more 
accurate than, most of the earlier spectroscopic studies\cite{explain}, and 
it agrees particularly well with the photoemission study\cite{gu}.  Thus 
all three direct methods (tunneling, optics and photoemission) yield 
very similar values.  

%As far as the 
%less direct measurements of the gap are concerned, let us first consider 
%the NMR data.
In early NMR measurements the absence of
the Hebel-Slichter (H-S) peak in the relaxation rate was interpreted
as an evidence for strong coupling.\cite{tycko,carbottenmr} More recently, 
Stenger {\it et al.} \cite{stenger} observed the H-S peak, and argued that 
Rb$_{3}$C$_{60}$ is in the weak coupling regime.  In particular, 
$\eta \approx 4.8$ was excluded as too high.  However, due to the many 
factors influencing the magnitude of the H-S peak, smaller values of 
$\eta$ could not be excluded.  Interestingly, Stenger {\it et al.} 
puts a lower bound of 200cm$^{-1}$ on the phonon frequency, and our result is 
entirely compatible with that constraint.  

Kiefl {\it et al.} determined $\eta \approx3.6 \pm 0.3$ from muon spin rotation 
experiments.  The interpretation of the results is similar to the NMR 
relaxation rate, except the low temperature limiting behavior is 
more accessible with muons. The spin relaxation rate depends on the 
number of electrons in the conduction band.  For BCS superconductors 
these electrons are thermally activated, thus the relaxation rates 
should exhibit Arrhenius temperature dependence at low $T$.  
This straightforward interpretation becomes rather complicated 
for intermediate and strong coupling.  At finite temperatures, the density of 
states is non-zero at the Fermi level (unlike the BCS limit).\cite{rainer}  
Indeed, Kiefl {\it et al.} used a broadened density of states, such as described 
in Eq. \ref{denstate}.  The value of the gap depends on the 
assumptions made for the temperature dependence of $\delta$. We believe that  
in this case it is crucial to use the full Eliashberg calculation to 
obtain a more reliable value for $\Delta$.  
Notice the contrast to the direct spectroscopic studies, where the 
phenomenological parameter
$\delta$ does not influence the numerical magnitude of the gap in any 
significant way.

The  dimensionless electron -- phonon coupling constant and the characteristic 
phonon frequency can be determined as follows.  According to Carbotte 
\cite{carbotte}, for various superconductors $\eta$ can be well approximated 
by 
\begin{equation}
\eta 
%= {2\Delta \over k_B T_{\rm c}} 
\approx 3.53 \left( 1+ 12.5 \left( {k_B 
T_{\rm c} \over \hbar \omega_0} 
\right)^2 \ln { \hbar \omega_0 \over 2 k_B T_{\rm c}} \right) .
\end{equation}
%Here $\omega_0$ is related to the the Eliashberg
%spectral density $\alpha^2 F$ by $ \ln \omega_0 = (2/\lambda) \int_0^\infty 
%\ln (\omega) \alpha^2 F/\omega \; {\rm d} \omega$ 
%and $\lambda = \int_0^\infty \alpha^2 F/ \omega \; {\rm d} 
%\omega$.\cite{allen,carbotte}  
The $\eta$ value determined here corresponds to $T_{\rm c}/ \omega_0  
= 0.10$.  This, in turn, leads to the dimensionless coupling constant 
of $\lambda = 0.89$ (from Eq. 2 of Allen and Dynes\cite{allen}) or a slightly 
larger value if the repulsive electron electron interaction is also 
considered (one obtains, for example, $\lambda = 1.16$ for $\mu  = 0.1$).  

Two important conclusions can be drawn from our results.  First, the 
coupling constant is reasonably close to the $\lambda_{tr} \approx 
0.65-0.80$, obtained from the DC electrical transport measurements by Vareka 
and Zettl\cite{vareka}.  Notice that for most BCS superconductors $\lambda$ 
and $\lambda_{tr}$ are close, but the two quantities do not have to be 
equal \cite{lambda}.   This agreement makes Rb$_{3}$C$_{60}$ look like one of the most 
simplistic materials.  We have to keep in mind, however, that with 
$\lambda_{tr} \approx 1$ the mean free path of the electrons is actually 
shorter than the lattice spacing. 
%(although it is longer than the C-C distance within a fullerene molecule).   
The existence of metallic 
conduction with such a short mean free path remains one of the mysteries of 
the transport in alkali fullerides. 

The second conclusion concerns the characteristic phonon frequency.  It 
turns out to be $\omega_0 \approx210$cm$^{-1}$, which is larger than the 
typical frequency of the acoustic phonons.   This result
is in agreement with the photoemission studies and calculations of Gunnarson 
{\it et al.}\cite{gunnarson},
where the low lying Raman modes of the C$_{60}$ molecule proved to contribute 
significantly to the electron -- phonon coupling.   
%suggest that the 
%electron -- electron coupling responsible for superconductivity is 
%mediated by one of the 
%low lying vibrational modes of the C$_{60}$ molecule\cite{martin}. Raman modes
%have the appropriate symmetry for electron -- phonon coupling; the lowest 
%lying $H_g(1)$ Raman active vibrational mode  of the pure C$_{60}$ is at 
%270cm$^{-1}$.  

In summary, we investigated the energy gap of the superconducting state of 
Rb$_3$C$_{60}$ and we demonstrated that this material 
cannot be described by the (weak coupling) BCS theory.  The dimensionless 
electron -- phonon coupling constant and the typical phonon frequency were 
determined.
It is expected that the Eliashberg theory, with a coupling constant in 
the order of 1, fully describes the superconductivity this material.  

\acknowledgments
We are indebted to P.B. Allen for enlightening discussions, to  G.P. Williams and G.L. Carr 
for valuable advice during the measurements at the NSLS, and to S. Lindaas and C. Jacobsen 
for the use of the scanning atomic force microscope.
This research has been supported by the Swiss National 
Science Foundation, by the NSF grants DMR-9501325 and INT-9414840, and by the 
Hungarian National Science Foundation grant OTKA-T015552.  
The NSLS and the U4IR beamline is 
supported by the US Department of energy, under the grant DEFG-0291-ER4531.

\end{document}